\documentclass[numreferences]{kluwer}
\usepackage[english]{babel}
\usepackage[dvips]{graphicx}

\newcommand{\beq}{\begin{equation}}
\newcommand{\eeq}{\end{equation}}
\newcommand{\bea}{\begin{eqnarray}}
\newcommand{\eea}{\end{eqnarray}}
\newcommand{\bml}{\begin{subequation}}
\newcommand{\eml}{\end{subequation}}

\input epsf

\begin{document}

\begin{opening}
\title{Modeling hydration water and its role in polymer folding}

\author{Pierpaolo \surname{Bruscolini}\thanks{E-mail address:
{\tt pbr@athena.polito.it}.}}
\author{Lapo \surname{Casetti}\thanks{Present address: INFM, UdR Firenze,
Dipartimento di Fisica, Universit\`a di Firenze, Largo Enrico Fermi 2, I-50125
Firenze, Italy. E-mail address: {\tt casetti@fi.infn.it}.}}

\institute{Istituto Nazionale per la Fisica della Materia (INFM) and
Dipartimento di Fisica, Politecnico di Torino, Corso Duca degli Abruzzi 24,
I-10129 Torino, Italy} 

\begin{abstract}
The hydrophobic effect is the dominant force which drives a protein towards its
native state, but its physics has not been thoroughly understood yet. We
introduce an exactly solvable model of the solvation of non-polar molecules in
water, which shows that the reduced number of allowed configurations of water
molecules when the solute is present is enough to give rise to hydrophobic
behaviour. We apply our model to a non-polar homopolymer in aqueous solution,
obtaining a clear evidence of both ``cold'' and ``warm'' collapse transitions
that recall those of proteins. Finally we show how the model can be adapted to
describe the solvation of aromatic and polar molecules.
\end{abstract}

\classification{PACS number(s)}{05.20.-y; 05.40.Fb; 61.25.Hq;
87.10.+e}

\end{opening}

\section{Introduction}

The hydrophobic effect, namely, the free energy cost that non-polar solutes
pay when transferred into water, is believed to be the dominant driving 
force of protein folding \cite{Dill}. 
In the native state of real proteins, in fact, 
non-polar residues are buried in the interior of the structure, thus minimizing
the exposure to water and the subsequent free energy cost.

Nonetheless, the physical properties of liquid water underlying the phenomenon
of hydrophobicity and giving rise to the characteristic behaviour of the change
of the thermodynamic functions upon solvation of a non-polar compound (not only
the free energy increase, but also the characteristic 
tem\-pe\-ra\-ture-de\-pen\-dence of the excess
specific heat, of the internal energy and of the entropy) 
are not yet completely understood, despite the extensive 
studies devoted to their investigation, ranging from simplified models
\cite{Muller} to numerical simulations (see
Ref.\ \cite{Dillea99} and references quoted therein). In many studies
hydrophobicity has been related to the ordering of water molecules around 
the solute \cite{Creibook,Dillea98}, but the question is still controversial, since this behavior, indeed
detected in simulations \cite{Dillea98},
could be a by-product and not the origin of hydrophobicity. An
alternative explanation suggests that  hydrophobic behaviour is
related to the process of opening a cavity in water to insert the nonpolar
solute, and to the interaction between water and solute
\cite{ArtHay98}. The decrease in entropy would not be related to
bond-induced ordering of water molecules, but to the opening of the
cavity.  

Here we discuss a model of hydrophobic solvation which is able to test 
the ``ordering'' hypothesis {\em directly}. 
First, we show that the smaller number of allowed configurations
of water molecules around a nonpolar compound
is enough to produce hydrophobic behaviour.  
Then we apply our model, which we keep as simple as to be analytically
integrable, to the case of polymer solvation, and recover for a
nonpolar homopolymer both ``cold'' and ``warm'' collapse
transitions that
recall those of proteins \cite{Privalov}, thus strengthening the idea, already
put forward in Refs.\ \cite{pl2000,svezia},
that an explicit, though simplified, description of water molecules around
non-polar solutes can provide a framework for a unified treatment of both the
``warm'' and the ``cold'' collapse transitions of polymers and proteins in
aqueous solution. We recall that most treatments of protein folding, where
water is not taken into account explicitly and hydrophobicity is described 
through effective potentials, do not allow a description of cold unfolding. 
Thus the model introduced here might well be relevant to the problem of 
protein folding. Finally we show how our model can be adapted to describe not
only the solvation of purely non-polar (aliphatic) molecules, but also of 
aromatic and polar ones. This opens the road towards a definition of a model of
the solvation of a real protein.

\section{Modeling hydration water}

Let us now introduce our model. Consider the water molecules belonging to 
the hydration  shell around  an isolated solute: the solute will affect
their geometric arrangement, causing a stronger spatial correlation than in
the bulk case. We describe water at the scale of the cluster of molecules 
which are spatially correlated in the presence of a nonpolar 
solute: interactions between clusters are not explicitly considered
and correlations are lost above this scale. 
This is consistent with the
observation \cite{Privalov} that the contributions to 
thermodynamic functions from the different 
chemical groups of a residue's side chain are to a good extent additive,
suggesting the existence of a length scale up to which molecules are spatially
correlated  due to  the solute, while correlation 
is small and can be neglected on larger scales.   

\begin{figure}
\epsfysize= 6 truecm 
\begin{center}
\mbox{\epsfbox{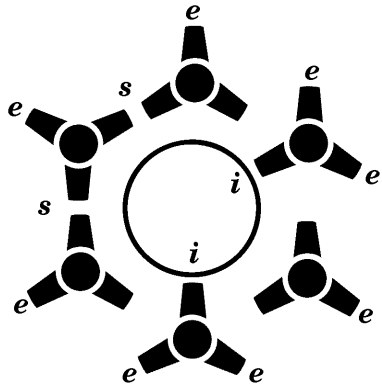}} 
\end{center}
\caption{Schematic picture of a hydration water cluster according 
to the model described in the text. The cluster molecules form bonds
that can be divided into
external $(e)$, on-shell $(s)$ and internal $(i)$ bonds.
In the external 
region  molecules of other clusters can be found, while 
the internal region contains either
a solute (shell case) or other water molecules (bulk case).
Bonds
can be broken (with probability $p_1$: see text) 
if their energy is larger than a given
threshold. 
}

\label{fig:bonds}
\end{figure}

%\begin{figure}
%\epsfysize= 6 truecm 
%\begin{center}
%\mbox{\epsfbox{fig_modello.eps}} 
%\end{center}
%\caption{Schematic picture of a hydration water shell according 
%to the model described
%in the text. The picture to the left shows the low-$T$ situation, 
%with all the bonds unbroken, while the one to the right depicts a high-$T$ 
%situation, 
%where some of the bonds are broken due to displacement of some
%molecules. 
%Bonds
%can be 
%broken also without displacement, if their energy is larger than a given
%threshold. The object in the interior of the shells can be either a solute
%(shell case) or one (or more) water molecules (bulk case).}
%\label{fig:cluster}
%\end{figure}

For the sake of simplicity we describe water molecules  as 
two-di\-men\-sio\-nal objects with three  hydrogen-bonding 
arms \cite{BellBenNaim,Dillea98}, representing the projection on a
plane of the tetrahedral coordination of a real water molecule. The
three arms are equivalent: no distinction is done between hydrogen
donors and acceptors.  
A schematic picture of a hydration water
shell according to our model is presented in Fig.\ \ref{fig:bonds}. 
Given a cluster of $m$ molecules, we 
assume that its ground state is characterized by a completely formed 
hydrogen-bond network, not only in the bulk case, but also when a
solute is present (this  involves geometric conditions on the
solute's shape and size, which we take for granted).
Thus each molecule in the ground state has $3/2$ hydrogen bonds,
but their energy can be different in the ``bulk'' and ``shell''
cases: namely, the energy difference per molecule, 
normalized to a bulk hydrogen-bond
energy $h_b$, is: 
\beq
K =   \frac{3}{2} \left(1 - 
 \frac{h_s}{h_b}\right) + J ~,
\eeq
where $h_s$ is the bond energy for shell molecules, and $J$ takes into
account all the contributions not related to hydrogen bonds (Van der
Waals and so on). Here $h_\bullet$ are positive quantities, while $J$
and $K$ can be positive or negative.

\begin{figure}
\epsfysize= 6 truecm 
\begin{center}
\mbox{\epsfbox{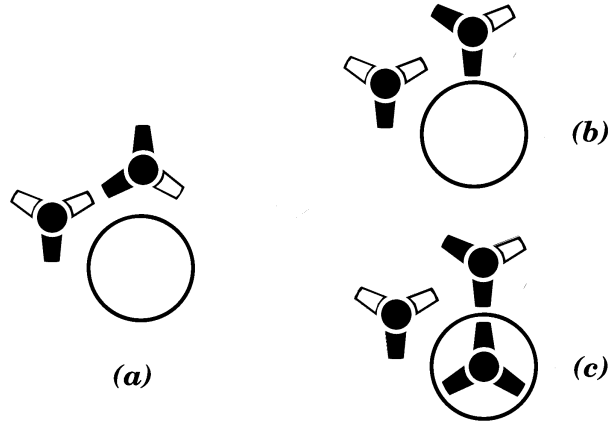}} 
\end{center}
\caption{
An example showing the dependence of the ``geometric'' contribution
to the density of states $g_\bullet$, $\gamma_\bullet$ 
in Eq.\ (\protect\ref{dos}), 
on the presence of the solute. 
To  keep the correct number of bonds per molecule we fictitiously
split water molecules in the cluster in two groups: 
molecules of group $a$ will be
able to form only one bond (one black arm), while two bonds will be
attributed to group $b$ ones.
Assuming that bonds can be broken independently,
in the situation $(a)$
depicted on the left,  where both molecules are in state $1$ (with two
on-shell and one external arm; see text),
every bond may be broken or formed. If we want
to evaluate the number of 
configurations allowing $\nu$ intact bonds, this
particular arrangement will be counted for every value of $\nu$
ranging from $0$ to $\frac{3}{2} m$ (three, in this case).
But if we rotate one of the molecules (right picture) to state $2$
(with one internal arm),  
the shell  bond is always broken for geometric reasons. The missing 
 bond is definitely lost in the shell case $(b)$, but might be recovered 
by the internal arm in the bulk water case $(c)$
This means 
that this configuration will  contribute to the $\nu=\frac{3}{2} m$ 
case  just in  bulk water. 
When the solute is present, at least one bond is broken, so
this configuration will contribute to the  
$\nu=\frac{3}{2} m - 1 $ case (1 bond
broken) down to the all-broken case $\nu=0$, but not to the completely
bonded  $\nu=\frac{3}{2} m$.  
This example helps one to understand
why $\gamma_s < \gamma_b$ when few bonds are broken.}
\label{fig:gamma}
\end{figure}

Our goal is to evaluate the partition function for the cluster in both
cases: 
\beq
{\cal Z}_{\bullet}^{{\rm clu}} = 
\int_0^\infty d\varepsilon \, g_{\bullet} (\varepsilon)\,
e^{-\beta {\cal H}_\bullet^{{\rm clu}} ({\varepsilon})}
\label{zclu}
\eeq
where $\bullet=b,s$ in the bulk and in the shell case, respectively, 
$g_{\bullet} (\varepsilon)$ is the density of states of the cluster at
the energy $E=h_b \varepsilon$ above its  ground state and ${\cal
H}_\bullet^{{\rm clu}}$ is
\beq
{\cal H}_\bullet^{{\rm clu}} = 
 h_b \left(  \varepsilon + K m \,
\delta_{\bullet, s} \right)~.
\label{hamclu}
\eeq
In the above framework, all the important features determining the
system thermodynamics are encoded in the density of states $g_{\bullet}
(\varepsilon)$ (rather than in the Hamiltonian, which has been substituted by
its value). To evaluate $g_{\bullet} (\varepsilon)$, we assume that each bond
in the system can be broken or formed 
independently of the others,  and that every formed one can be described by
a harmonic potential. A bond is broken if its energy\footnote{Considering
independent bonds allows one to speak of the ``energy of a bond'', even if in
principle one can only speak of a mean energy per molecule.} exceeds 
$h_{\bullet}$, while to 
form a bond one needs also to have a bonding partner whose arms
are correctly oriented. In other words, all the configurations with a given
number of broken bonds are degenerate in energy, and this degeneracy depends on
geometrical constraints, which can be different in the bulk and shell cases.
Thus, at a given energy $\varepsilon$ there will be a probability 
$p_{\bullet}(\lambda,\varepsilon) $ to break $\lambda$ bonds, and
a given number of broken  bonds can be obtained with
a set of configurations  of water  molecules whose number depends on
the presence or absence of a nonbonding solute (see Fig.\ \ref{fig:gamma}). The
important point is that when  few bonds are broken 
the available number of configurations is {\em
smaller} in the shell case rather in the bulk one, i.e., the solute indeed
forces an ``ordering'' of the water cluster. 
We shall see in the following that this fact, together with a
different value
for $h_s$ and $h_b$,
is enough to get hydrophobic behaviour.

Hence, a reasonable ansatz  for the functional form of $g_{\bullet}
(\varepsilon)$ appears to be:
\beq
g_{\bullet}(\varepsilon) = 
\sum_{\lambda = 0}^{\frac{3}{2}m} p_{\bullet}(\lambda,\varepsilon)\,
\omega(\lambda,\varepsilon) \, 
\gamma_{\bullet} \left(\frac{3}{2} m - \lambda\right)~, 
\label{dos}
\eeq
where 
$\omega(\lambda,\varepsilon)$ is the density of states of the system
of harmonic oscillators resulting when $\lambda$ bonds are  broken,
and $\gamma_{\bullet} \left( \nu \right)$ is the number of geometric
arrangements of the water molecules allowing $\nu$ unbroken bonds.
Assuming independence of the bonds, we can write: 
\beq
%p_{\bullet}(\lambda,E) = \tbinom{\frac{3}{2} m}{\lambda} 
p_{\bullet}(\lambda,\varepsilon) = {\frac{3}{2} m \choose \lambda} 
p_1^\lambda \left( 1 - p_1\right)^{\frac{3}{2} m - \lambda} 
\label{plam}
\eeq
where $p_1$ is the probability that a bond acquires an energy larger
than  $h_\bullet$ and breaks, when the cluster energy $h_b \varepsilon$ is
equipartitioned  on $D$ degrees of freedom, so that
$p_1 = e^{-h_{\bullet}/\tilde{T}}$ and  
$\tilde{T} = \frac{2 \varepsilon h_b}{D}$.
Notice that in principle $D = D(m,\lambda)$  also depends on $\lambda$, as
well as $\omega(\lambda,\varepsilon)$ does: the density of states of
interacting 
molecules depends on how many molecules are
bonded and how many are free, which in turn is a function of $\lambda$.
Yet, for the sake of simplicity, we assume the  degrees of freedom to
be always those of bonded molecules, so that $D= 2 m f$ and
\beq
\omega(\lambda,\varepsilon) \simeq \omega(\varepsilon) =
{\cal C} \, \varepsilon^{\frac{D}{2} - 1}~,
\label{omega}
\eeq
where $f$ are the degrees of freedom of one molecule ($f=3$ for the 
2-dimensional case), and ${\cal C}$ is a constant. 

To estimate $\gamma_{\bullet} \left( \nu \right)$ in the 2-dimensional case we
consider a cluster and  divide the space in three regions: the
cluster's molecules are aligned in the shell region, that separates
the external one, where water molecules of other clusters are
found, from the internal one, which contains either the nonbonding solute or
other water molecules in the ``s'' or ``b'' cases, respectively. 
Thus, we talk about  ``external'', ``on-shell'' and
``internal'' bonds, according to the region they point towards (see
 Fig.\ \ref{fig:bonds}). Moreover, we classify the orientation
 of a molecule of the cluster according to the direction of its arms,
 considering, for the sake of simplicity, only two states: state 1, with two
 on-shell arms and the third pointing outside, and state 2, with two
 external and one internal arms. In the latter, the internal bond
 will always be broken when the solute is present, while in the bulk water case
 it can be formed with a probability accounting for geometric restrictions 
 on  the orientation of the internal water. 
 We assume no restriction on external bonds:
they can always be  formed. In this framework, $\gamma_{\bullet} \left( \nu
\right)$ is related to the number of arrangements of the $m$ molecules
in the two  states allowing for $\nu$ bonds to be formed, and in the end one
gets: 
\beq
\gamma_{\bullet} \left( \nu \right) = 
\sum_{\nu_s=0}^{\nu} \sum_{\nu_e=0}^{\nu-\nu_s}  
\delta_\bullet \left( \nu_e,\nu_s,\nu_i \right)
\, , 
\label{gamma}
\eeq
where  $\nu_i = \nu- \nu_s- \nu_e$,
\beq
\delta_\bullet \left( \nu_e,\nu_s,\nu_i \right)\!\! =
\sum_{s=0}^{m}\sum_{k=0}^{k_{{\rm max}}}  \sum_{j=0}^{s} 
\sum_{i=0}^{j}   
\xi_\bullet \left( s, k, j, i,  \nu_e, \nu_s,\nu_i \right) \, ,
\label{delta}
\eeq
with $k_{{\rm max}}=\min(s, m-s-\nu_s)$ and
\bea
&&\lefteqn{\xi_\bullet \left( s, k, j, i, \nu_e,\nu_s,\nu_i \right) =  
\pi (s, k) \pi_{{\rm h}} \left(j; s,\frac{m}{2},m \right) 
\pi_{{\rm h}} (i; j, q_\bullet, m)  \times} \nonumber \\ 
&&
\,\pi_{{\rm b}} \left(\nu_i, i,\frac{1}{2} \right)  
\pi_{{\rm b}} \left(\nu_e, \frac{m}{2}+s-j,\frac{1}{2} \right)
\pi_{{\rm b}} \left(\nu_s, m-s-k,\frac{1}{2} \right)\,.
\label{xi}
\eea
In the above equation the following definitions hold:
\bml
\beq
\pi (s, k) = \frac{1}{2^{m}}
\left[{s \choose k}{m-s \choose k}(1-\delta_{s m}) +
\delta_{s m}\delta_{k 0}  \right],
\label{pi_plain}
\eeq
\beq
\pi_{{\rm h}} (s; n, S, N) = {S \choose s} {N-S \choose n-s} 
\left[{N \choose n}\right]^{-1}\,,
\label{pi_h}
\eeq
\beq
\pi_{{\rm b}} (j, n,p) = {n \choose j} p^j (1-p)^{(n-j)} \,.
\label{pi_b}
\eeq
\label{pi}
\eml
Let us now explain how these
expressions can be derived. 
To keep the right number of bonds per molecule (i.e., 3/2), let us 
fictitiously 
split the $m$ molecules into two groups of $m/2$ each.
Group {\em a}-molecules will be able to form one bond: 
a shell one when in state 1 
and an external one when in state 2.
Group {\em b}-molecules will be able to form two bonds: a shell one 
and the external one in state 1, 
an external and the internal one in state 2 (See. Fig.\ \ref{fig:gamma}). 
Hence, state 1 of both groups 
is able to form shell bonds, while internal bonds are possible only in the 
$(b,2)$ state. Given the number $s$ of 
molecules in the state 2 (not shell-bonding) and 
the number $k$ of groups of state 1-molecules between state 2-ones, 
the probability of making
$\chi = m-s-k$ shell bonds is 
$\pi (s,k)$ given in Eq.\ (\ref{pi_plain}).
The probability that one of these configurations also has  
$i$ internal bonds depends, first of all, on the probability of 
fishing out $j$ $(b,2)$-molecules among the total of $s$ in state 2, 
given the total number of molecules $m$ and the total number of $b$-molecules 
$m/2$, so that one gets a hypergeometric probability -- defined in 
Eq.\ (\ref{pi_h}) -- $\pi_{{\rm h}} (j; s,\frac{m}{2},m)$. 
This is not enough, though, because, for geometrical
reasons, some of the internal arms will not find a bonding partner.
Indeed, we assume that there are just $q_\bullet$ (out of $m$) positions where 
internal bonds may actually be formed. Their number distinguishes the bulk from
the shell-water case: $q_s=0$ with a  non-bonding solute, 
while in the bulk case $0< q_b \le m$. Again we have a 
hypergeometric probability of placing $i$ of the $j$ 
molecules with internal arms, in the $q_\bullet$ good positions 
for bond formations, on a total of $m$ possibilities: 
$\pi_{{\rm h}} (i; j, q_\bullet, m)$. 
The product of the above probabilities gives the fraction 
of the total number of conformations
that is able to form $\chi$ shell bonds, $i$ internal ones and $m/2 +s-j$ 
external ones, assuming that all the molecules with external arms -- i.e., group
$b$ and state $(a,2)$ -- form external bonds.
If we now let the bonds be also broken, any geometric 
arrangement of the $m$ molecules allowing $\nu$ bonds also contributes
to the cases where fewer bonds are formed. Indeed, given a geometric 
arrangement, we can choose to keep or break the external, internal and 
shell bonds with the binomial probability
$\pi_{{\rm b}}$ defined in Eq.\ (\ref{pi_b}), 
whence the expression of $\xi_\bullet$ in Eq.\ (\ref{xi}).
Upon summing over all the geometric arrangements that can contribute to a 
pattern with $\nu$ bonds, we obtain the expression (\ref{gamma}) for 
$\gamma_{\bullet} \left( \nu \right)$. 

Upon substituting Eqs.\ (\ref{plam},\ref{omega},\ref{gamma}) 
into Eq.\ (\ref{dos}) and then into Eq.\ (\ref{zclu}), we get:
\beq
{\cal Z}_{\bullet}^{{\rm clu}} = 
A_\bullet \, e^{-\beta h_b K m  \delta_{\bullet,s}}
\label{zclu2}
\eeq
where
\beq 
A_\bullet ={\cal C} 
\sum_{\lambda = 0}^{\frac{3}{2} m} {\frac{3}{2} m \choose \lambda}
\gamma_{\bullet} \left(\frac{3}{2} m - \lambda\right) 
I_\bullet \left( \lambda \right)\, ,
\eeq
${\cal C}$ is a constant and 
\beq
I_\bullet \left( \lambda \right) = 
\frac{\left(  D-1 \right)!}{\left(\beta h_b \right)^{D}} 
\delta_{\lambda,0} + \sum_{j=1-\delta_{\lambda,0}}^{3 m/2 -\lambda} 
{\frac{3 m}{2} -  \lambda \choose j }(-1)^j 2 
\sigma_\bullet^{D/2} K_{D}(\tau_\bullet) \,,  
\eeq
where $K_n(z)$ is the Bessel-$K$ function and the following definitions hold:
\beq
\sigma_{\bullet}=\frac{h_\bullet D \left(j+\lambda \right)}{\beta h_b^2}\, ,
\eeq
\beq
\tau_{\bullet}=2 \sqrt{ \beta h_\bullet D \left(j+\lambda \right)}\,.
\eeq

\section{Hydrophobic solvation}

Having defined a model for hydration water which should take into account the
main physical ingredients relevant to the thermodynamics of solvation, let us
now apply our model to two physically interesting cases which can be dealt
with in a purely analytical way. To start with, we
consider the solvation of a single non-polar compound, then we apply our model
to a non-polar homopolymer.

\subsection{Transfer of a nonpolar solute into water}

\begin{figure}
\epsfysize= 5 truecm 
\begin{center}
\mbox{\epsfbox{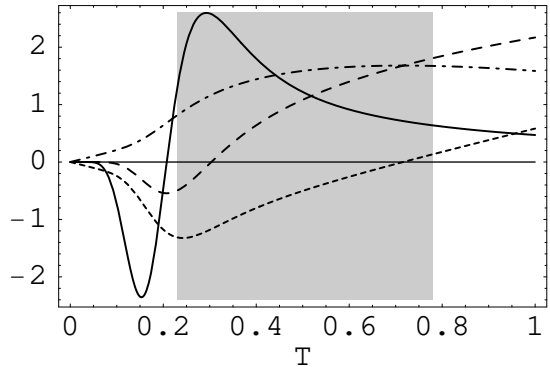}} 
\end{center}
\caption{Free energy, energy, entropy and specific heat 
changes upon solvation of a nonpolar solute:
 $\Delta F$ (dash-dotted line), $\Delta E$, (dashed), $T \Delta S$ (dotted), 
$\Delta C/3$ (solid line; $1/3$ is for rendering purposes). 
Here $m=4$, $q_b=m/2$ (half of the molecules can form internal bonds,
in the bulk case), $h_s/h_b =1.2$, $J=0.3$
(so that the ground-state energy shift $K=0$). Energies and temperatures are 
expressed in bulk hydrogen bonds  units 
($h_b$); Boltzmann constant $k_B$ is set to 1 and 
specific heat is, accordingly, adimensional. The shaded region can be compared
with numerical simulations (e.g., with  Fig.\ 5 of 
Ref.\ \protect\cite{Dillea98}) and experimental results like those reported in
Ref.\ \protect\cite{Privalov}.}
\label{fig:solutetherm}
\end{figure}

Let us  consider the cluster of $m$ water molecules 
in the bulk case and substitute the internal water  with a  nonpolar solute. 
This describes the transfer of a nonpolar molecule from its gas 
(non-interacting) phase to water, i.e., hydrophobic solvation. We study 
the difference in thermodynamic functions. The free energy change is:
\beq
\Delta F = F_{{\rm solution}}- F_{{\rm water}} = - \frac{z}{\beta}\log{x}\,,
\eeq
where 
\beq
x=  \frac{A_{s}}{A_{b}}\, 
e^{- \beta K h_b m  } \,.
\label{formulazze}
\eeq
Similar expressions hold for  energy, entropy and specific heat changes.
%\bml
%\beq
%\Delta E = E_{{\rm solution}}- E_{{\rm water}} = ?\,;
%\eeq
%\beq
%\Delta S = S_{{\rm solution}}- S_{{\rm water}} = ?\,;
%\eeq
%\beq
%\Delta C = C_{{\rm solution}}- C_{{\rm water}} = ?\,.
%\eeq
%\eml
The temperature dependence of these functions, 
reported in Fig.\ \ref{fig:solutetherm}, shows the hallmarks 
of hydrophobic behaviour, even for $K=0$ (i.e., without any 
ground-state energy difference). 
In fact, we find a maximum in the free energy
cost, a pronounced and positive peak in the specific heat difference $\Delta
C$, and minima  in both $\Delta E$ and $T \Delta S$; then, as $T$ grows,
$\Delta E$ and $T \Delta S$ cross the zero -- thus defining the characteristic
temperatures commonly referred to as $T_H$ and $T_S$ -- 
and eventually become positive.
Notice that the peak in $\Delta C$ cannot be recovered within the
Information Theory Approximation, usually applied together with the
cavity approach \cite{ArtHay99}. 

\subsection{Polymer in solution}

Let us now turn to the study of a nonpolar 
homopolymer in solution, taking into account 
just the behaviour of water clusters in  the vicinity 
of a monomer, and disregarding interactions between monomers 
and  between water clusters. Our goal is, in fact, to understand 
the effect of the hydrophobic effect alone on polymer behaviour.

We model a polymer as a $N$-step 
self-avoiding walk on a two dimensional lattice with coordination $z$.
On each lattice site there can be
either a monomer or $z$ clusters of $m$ water molecules, so that each 
monomer-water contact involves one cluster. 
The Hamiltonian follows from Eq.\ (\ref{hamclu}):
\beq
{\cal H} = \sum_{j=1}^{N_W} h_b \left( \sum_{\mu=1}^{z} 
\varepsilon_{j \mu} \, + K m l_j \right) \,,
\label{hampoly}
\eeq
where $N_W= (z-2) N + 2$ is the  highest number of water sites that can 
be in contact with the polymer, and $l_j$ is the number of contacts 
between the $j$-th water site and the monomers.
The partition function of the polymer in solution reads as:
\beq
{\cal Z} = \sum_{C} Z(C) = \sum_{n_c}
\zeta \left(n_c \right) Z (n_c)
 \eeq
where $C$ are the conformations of the polymer and $Z(C)$ 
the restricted partition function, obtained tracing over water variables
at fixed conformation $C$. Due to the form of the Hamiltonian given in 
Eq.\ (\ref{hampoly}),  $Z(C)$ depends only on the total number of 
water-monomer contacts $n_c$, and $\zeta \left(n_c \right)$ is the 
number of  SAWs  characterized by the same value of $n_c$.
$\cal Z$ can be factorized as
\beq
{\cal Z} = {\cal Z}_b {\cal Z}_I\,,
\eeq
where ${\cal Z}_b =({\cal Z}_b^{{\rm clu}})^{ z N_W}= 
A_b^{z N_W}$ is the contribution of all water sites when in
contact to other water, and  
\beq
{\cal Z}_I = \sum_{n_c} \zeta \left( n_c \right) x^{n_c}\,. 
\eeq
In the following we shall study the specific heat of the system, that,
 according to the above factorization, is the sum of a bulk contribution
$C_b$ and of an interaction contribution $C_I$. We shall also study 
the average number of  water-monomer contacts:
\beq
\langle n_c \rangle = x \frac{\partial}{\partial x } \log{ {\cal Z}_I} \,,
\eeq
which is a measure of the compactness of the polymer.
To exactly evaluate the above  quantities, an exhaustive enumeration of 
the SAWs should be performed, in order to obtain $ \zeta \left( n_c \right)$.
However, if we restrict ourselves to a square lattice, the numerical
estimates reported in \cite{Dougea} allow us to write
\beq
\zeta \left( n_c \right) \simeq \zeta_0 \frac{1}{w(n_c)!} 
(\alpha_0 N)^{w(n_c)} \exp(\alpha_0 N)~,
\eeq
where $w(n_c)=(N_W-n_c)/2$ is the number of monomer-monomer contacts,
$\zeta_0$ is the number of SAWs of length $N$ and $\alpha_0 = 0.164$.
Such an
estimate is expected to be very good if $N$ is large \cite{Dougea}.
Hence, an analytical expression can be found for ${\cal Z}_I$:
\beq
{\cal Z}_I=\frac{1}{w_{{\rm max}}!}x^{N_W} 
\Gamma (w_{{\rm max}}+1, \frac{\alpha_0 N}{x^2}) 
e^{  -\alpha_0 N \left( 1- \frac{1}{x^2} \right) } \,,
\label {zian}
\eeq
where $\Gamma(n,x)$ is the incomplete $\Gamma$-function and 
$ w_{{\rm max} } =(N_W - n_c^{{\rm min} })/2 $
($n_c^{{\rm min}}$ accounts for the fact that the smallest 
number of water-monomer contacts is determined by the length 
of the perimeter of the globule: $n_c^{{\rm min}} \simeq 2 \sqrt{N \pi}$).

\begin{figure}
\epsfysize= 5 truecm 
\begin{center}
\mbox{\epsfbox{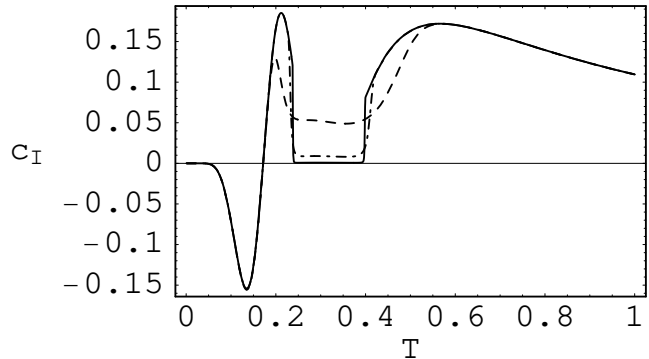}} 
\end{center}
\caption{Excess specific heat  $C_I/N_W$ of the nonpolar homopolymer 
in solution, for different lengths: $N=2\times10^2$ (dashed), $N=10^4$
(dash-dotted), $N=10^6$ (solid line).
Parameters as before.
}
\label{fig:cspec}
\end{figure}

\begin{figure}
\epsfysize= 5 truecm 
\begin{center}
\mbox{\epsfbox{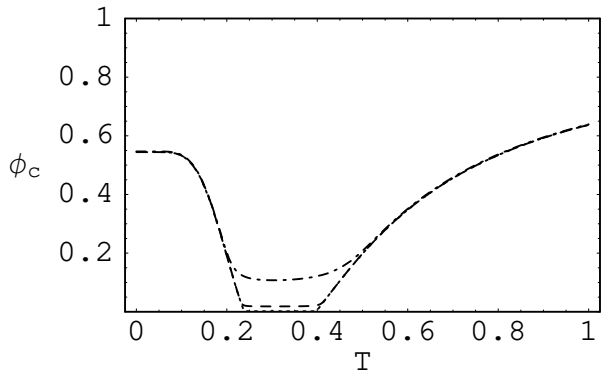}} 
\end{center}
\caption{Normalized average number of water-monomer contacts,
$f_c= \langle n_c \rangle/N_W$, for different polymer lengths 
(the same as before). Parameters and rendering conventions as before.}
\label{fig:nc}
\end{figure}

The results obtained $C_I$ and  $\langle n_c \rangle$ stemming 
from Eq.\ (\ref{zian}) are reported in Figs.\ \ref{fig:cspec} and
\ref{fig:nc}, for different lengths of the polymer. 

The presence of both ``cold'' and ``warm'' collapse transitions, signalled by
the drop of the number of contacts and by the jumps in the specific heat is 
strikingly evident from our results. As $N$ grows both transitions get sharper,
thus suggesting the existence of a true phase transition in the thermodynamic
limit. This phenomenology is very close to that of proteins in solution: our
results confirm that, as
already suggested in Refs.\ \cite{pl2000,svezia}, the explicit treatment of
the solvent -- though in an extremely simplified way -- is a natural and
powerful way
to get a unified modeling of both the cold and the warm unfolding transitions 
in proteins. Moreover, it is worth noticing that a cold collapse qualitatively
very similar to that predicted by our model has been experimentally observed
also in homopolymers like the poly(N-isopropylacrylamide) \cite{coldexp}.

\section{Solvation of aromatic and polar compounds}

Proteins are heteropolymers whose monomers are aminoacids. The aminoacid 
residues (i.e., those parts of the aminoacids which can be exposed to water
once they are linked with other aminoacids in a
protein chain) can be roughly split into two
classes: polar (or even charged) 
and non-polar. However, as far as the qualitative features of
their interaction with water is concerned, not all the non-polar molecules 
behave in the same way. One can identify two families, aliphatic and aromatic
residues \cite{Privalov}, with different thermodynamic behavior. 
Aliphatic compounds are ``purely'' hydrophobic, i.e., show 
a positive jump in both the specific heat and the free energy upon solvation. 
On the
contrary, when an aromatic residue is transferred into water from the gas phase
there is a positive jump in the specific heat but there is a free energy gain
too,
i.e., $\Delta F < 0$. Aromatic molecules show, in a way, an intermediate
behaviour betweeen
hydrophobic and hydrophilic. Moreover, also in the case of
polar molecules there is a change in the thermodynamic functions: $\Delta F <
0$ and $\Delta C < 0$, so that they can be reasonably called hydrophilic.

The model of solvation discussed so far works very well when applied to purely
hydrophobic solvation, i.e., to the solvation of aliphatic monomers or
homopolymers. However, in view of a generalization of the approach presented here
to heteropolymers and especially to proteins, one should devise a way to model
also the solvation of aromatic and polar residues. As we are going to show in
the following, our model is able to accomplish both tasks, provided one chooses
the parameters according to the physical requirements of these situations. This
is not only very promising for future applications to proteins, but also tells
us that our model grasps, though in a very simplified way, some of the basic
physics of water-solute interaction.

\subsection{Aromatic molecules}

\begin{figure}
\epsfysize= 5 truecm 
\begin{center}
\mbox{\epsfbox{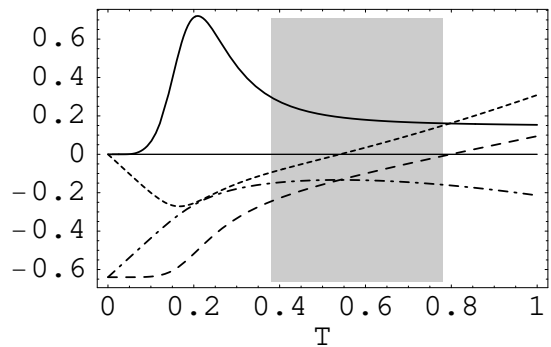}} 
\end{center}
\caption{Free energy, energy, entropy and specific heat 
changes upon solvation of an aromatic solute: rendering conventions as in 
Fig.\ \protect\ref{fig:solutetherm}. Here $m=4$, $q_b=m/2$, $h_s/h_b =0.98$, 
$J=-0.07$. The shaded region can be compared with experiments reported in 
Ref.\ \protect\cite{Privalov}.}
\label{fig:aromatic}
\end{figure}

Figure \ref{fig:aromatic} presents trends in the 
changes in thermodynamic functions upon solvation that 
qualitatively recall those obtained experimentally for  aromatic
residues \cite{Privalov}. The plots are obtained with $h_s/h_b = 0.98$, 
$J=-0.07$, $q_s=0$, $q_b=m/2$; more in general, $h_s/h_b \simeq 1$
and $q_b=m/4$ would provide
similar results, as far as the experimental  temperature window is concerned 
 (data not shown). In the framework of the model, this
implies that 
hydrogen bonds for bulk and hydration water are more or less
equivalent, and that shell molecules are less likely to form internal
bonds  than in the bulk case (or cannot at all). At the same time
there is a small gain, related to $J$, in hydrating the solute,
 suggesting that other interactions than hydrogen bond
could favour the aromatic hydration.
A clear intepretation of this fact is not straightforward: it is
unlikely that a negative $J$ could come  from Van der Waals
contributions prevailing over the electrostatics of bulk water:
electrostatics interactions between solute and solvent 
should come into play, instead.
It is difficult to say if their effect could be described by an
isotropic term $J$; we cannot exclude also that a
negative $J$ compensates the fact that in the model internal bonds have
the same energy as external and on-shell.
Even if the precise scenario could be determined only by all-atoms
simulation, the hypothesis that electrostatic interactions
between solute and water are more important for aromatic than for
aliphatic compounds is supported by the experimental evidence that
aromatic residues  present a partial polar nature, with an excess
negative charge on the faces of the aromatic ring and a partial
positive charge of the hydrogen atoms at the border \cite{Privalov,
BuPe88,Creibook}.  
  
\subsection{Polar molecules}
\begin{figure}
\epsfysize= 5 truecm 
\begin{center}
\mbox{\epsfbox{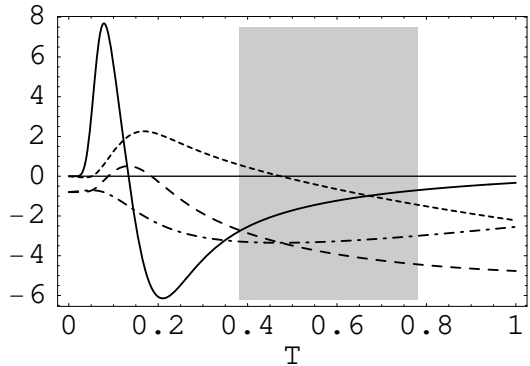}} 
\end{center}
\caption{As in Fig.\ \protect\ref{fig:aromatic} for a polar solute. 
Here $m=4$, $q_b=m/2$, $h_s/h_b =0.4$, $J=-0.95$.}
\label{fig:polar}
\end{figure}

Figure \ref{fig:polar} presents thermodynamic trends which are in  
a good qualitative agreement with  experimental  findings  for polar
residues  \cite{Privalov}. Again, the choice of the parameters is not
unique, since the region to be compared to experiments is not as
sensitive as the low T one to parameters changes, and a comparison of
a two-dimensional  model to real experiments cannot be made truly
quantitative. Nevertheless, there are some crucial aspects
characterizing polar behavior: a low value of $h_s/h_b$ (i.e. less
than 0.6) is required 
to produce the correct trend in $\Delta C$
together with a low value
for $T_s$, whence the relation $\Delta E<\Delta F$.  
On the other hand, a low ratio $h_s/h_b$ would yield a positive energy
shift $K$; thus, it
must be compensated by a large and negative $J$, favouring solvation, 
if the experimental values for a large and negative $\Delta F$ are to
be recovered.
It is important to notice that, for low values of $h_s/h_b$, 
the different degeneracies $ g_b (\varepsilon)$,  $ g_s (\varepsilon)$
in bulk and shell
water come essentially from the different probabilities of breaking
bonds: the geometric arrangement of the molecules almost plays no role.  

The need for a low ratio $h_s/h_b$ and $J < 0$ suggests a possible
physical explanation of the parameters: the
presence of a polar compound hinders the formation of hydrogen bonds
between water molecules by competing with it; a strong direct
solute-water interaction ($J$ is almost equal to $h_b$) results in
weakened hydrogen bonds among shell water molecules.

\section{Concluding remarks and future developments}
 
We have introduced and discussed a statistical-mechanical 
model of non-polar solvation in water. In our model the water degrees of
freedom are explicitly considered, though in a very simplified way: water
molecules are modeled as planar objects with three equivalent bonding
arms (roughly speaking, the projection of water's tetrahedral coordination
into a plane)
and
all the treatment is for a two-dimensional system. Nonetheless, this allows us
to obtain an exact analytic solution of the model, which unambigously shows
that hydrophobic behaviour can be related to a combination of the 
strengthening of hydrogen bonds together with the reduction 
of available configurations for water molecules when the
solute is present. 
Our model thus supports the conjecture that the  physical
origin of hydrophobicity is the formation of ``ordered'' cages of water
around the solute, with stronger hydrogen bonds than in bulk, 
even if we cannot say which one of
the two ingredients above is more determinant. 
Applying our model to a hydrophobic homopolymer we have
clearly shown that it is swollen at low temperatures, then becomes maximally
compact and eventually unfolds again at higher temperatures. This behaviour
closely reminds that of  proteins, which undergo both a cold and a warm
unfolding transition.

In view of the application of our model to proteins, we have shown that it can
be adapted to describe also the solvation of aromatic and polar molecules. To
study the behaviour of a model of a real protein along these lines, however, it
will be necessary to extend the model to the three-dimensional case and to
determine, by means of a comparison with experiments, the right parameters for
all the aminoacids. Work is in progress in these directions. 

\begin{acknowledgements}
We would like to thank all the participants of the workshop {\em Protein
folding: simple models and experiments} for providing a very stimulating
atmosphere. We would like to thank especially P.\ De Los Rios for many useful
discussions.
\end{acknowledgements}

\end{document}